\documentclass[a4paper,12pt]{article}
\usepackage{jcappub} % for details on the use of the package, please
\usepackage{hyperref}
\hypersetup{
    bookmarks=true,         % show bookmarks bar?
    unicode=false,          % non-Latin characters in Acrobat’s bookmarks
    pdftoolbar=true,        % show Acrobat’s toolbar?
    pdfmenubar=true,        % show Acrobat’s menu?
    pdffitwindow=false,     % window fit to page when opened
    pdfstartview={FitH},    % fits the width of the page to the window
    pdftitle={My title},    % title
    pdfauthor={Author},     % author
    pdfsubject={Subject},   % subject of the document
    pdfcreator={Creator},   % creator of the document
    pdfproducer={Producer}, % producer of the document
    pdfkeywords={keyword1} {key2} {key3}, % list of keywords
    pdfnewwindow=true,      % links in new window
    colorlinks=true,       % false: boxed links; true: colored links
    linkcolor=red,          % color of internal links
    citecolor=cyan,        % color of links to bibliography
    filecolor=magenta,      % color of file links
    urlcolor=blue,           % color of external links
    linktocpage=true
}
\usepackage{subfigure}
\usepackage{pstricks}
\usepackage{color}
\usepackage{amsfonts}
\usepackage{mathrsfs}
\usepackage{epsfig}
\usepackage{amsmath,amssymb,amsthm,graphicx,latexsym}
\usepackage{ulem}

\newcommand{\be}{\begin{equation}}
\newcommand{\ee}{\end{equation}}
\newcommand{\bea}{\begin{eqnarray}}
\newcommand{\eea}{\end{eqnarray}}
\newcommand{\bml}{\begin{subequations}}
\newcommand{\eml}{\end{subequations}}
\newcommand{\bfig}{\begin{figure}}
\newcommand{\efig}{\end{figure}}

\newcommand{\slashed}{\hspace{-1.1ex}/}

\title{%\boldmath \bf 
  An accurate bound on tensor-to-scalar ratio  and the scale of inflation}

\author[a]{Sayantan Choudhury}
\author[b]{ Anupam Mazumdar}

\affiliation[a]{Physics and Applied Mathematics Unit, Indian Statistical Institute, 203 B.T. Road, Kolkata 700 108, INDIA}
\affiliation[b]{Consortium for Fundamental Physics, Physics Department, Lancaster University, LA1 4YB, UK}

\abstract{In this paper we provide an accurate bound on primordial gravitational waves, i.e.  tensor-to-scalar ratio $(r)$ for a general 
class of single-field models of inflation where inflation occurs always below the Planck scale, and the field displacement during inflation remains sub-Planckian. If inflation has
 to make connection with the real particle physics framework then
 it must be explained within an effective field theory description where it can be trustable below the UV cut-off of the scale of gravity. We 
 provide an analytical estimation and estimate the largest possible  $r$, i.e. $r\leq 0.12$, for the field displacement less than the Planck cut-off.}

\begin{document} 
\maketitle
\flushbottom

\section{Introduction}

If the primordial inflation~\cite{Ade:2013uln} has to make connection to the observed world and be a {\it predictive science} then it has to be embedded within a particle 
theory~\cite{Mazumdar:2010sa}, where the last $50-60$ e-foldings of inflation must occur within a visible sector with a laboratory measured  inflaton couplings to the
 Standard Model physics in order to create the right form of matter with the right abundance~\cite{Allahverdi:2006iq}~\footnote{Note that after Planck there is no trace
 of isocurvature perturbations and there is a severe constraint on dark radiation~\cite{Ade:2013zuv}. Therefore the inflaton vacuum cannot be arbitrary as first pointed
 out in this review~\cite{Mazumdar:2010sa}. Models of inflation based on gauge invariant flat directions of Standard Model quarks and leptons naturally provides an ideal
 inflaton candidate embedded within a visible sector with well-known interactions, discussed in Ref.~\cite{Allahverdi:2006iq}, and their recent update after Planck~\cite{Wang:2013hva,Choudhury:2013jya}.}.
 This inevitably puts constraint on the vev of inflation, i.e. $\phi_0$, and the range of flatness of the potential during the observed $17$ e-foldings of 
inflation, i.e. $\Delta \phi$. For the simplest single field dominated model of inflation, there are two important constraints which all the models must satisfy.

\begin{itemize}

\item{$\phi_0\ll M_p$ - vev of the inflaton must be bounded by the cut-off of the particle theory,  where $M_p=2.4\times 10^{18}$~GeV. We are assuming that $4$ dimensions $M_p$ puts a natural cut-off here for any physics beyond the Standard Model.}

\item{ $|\Delta \phi | \ll M_p$ - the inflaton potential has to be flat enough  during which a successful
inflation can occur. Note that the flatness of the potential has to be fine tuned - there is no particle physics symmetry which can maintain the
flatness~\cite{Mazumdar:2010sa}. We will assume $V''(\phi_0)\approx 0$, where $V(\phi)$ denotes the inflaton potential, and prime denotes derivative w.r.t. the $\phi$ field.}

\end{itemize}

The aim of this paper is to impose these two conditions to obtain an improved bound on the tensor-to-scalar ratio, $r$.
In Refs.~\cite{BenDayan:2009kv,Hotchkiss:2011gz,Shafi:2010jr,Rehman:2010wm,Okada:2011en,Civiletti:2011qg}, it was realized that it is
 possible to obtain large $r\sim {\cal O}(10^{-1}-10^{-2})$
for small field excursion characterized by, $\phi_0\leq M_p$ and $\Delta\phi\leq M_p$. The bound on $'r'$ was further improved in the recent work, see Ref.~\cite{Choudhury:2013jya},
where it was demonstrated that it is possible to saturate the Planck limit on tensor-to-scalar ratio,
 i.e. $r\leq 0.12$~\cite{Ade:2013uln}.
 %~\footnote{We are also assuming that the initial conditions for the gravitational waves are
 %set by quantum initial condition. For a classical gravitational wave the amplitude of the tensor modes will be absolutely
% unobservable~\cite{Ashoorioon:2012kh}.}. 
 
Such a large tensor-to-scalar ratio can be obtained provided one deviates from a
monotonic behavior of slow roll parameter $\epsilon_V$, which we will elaborately discuss below by incorporating the effects of higher order
 slow-roll corrections for generic class of sub-Plackian inflationary models in presence of a non-negligible and scale-dependent running
of the scalar and tensor power spectrum \cite{BenDayan:2009kv,Hotchkiss:2011gz,Choudhury:2013jya,Easther:2006tv}.

In this respect we are improving on previously obtained  bound on large $r$, i.e. $r\sim {\cal O}(0.1)$, where $\phi_0$ and $|\Delta\phi | $ were taken beyond $M_p$, see~\cite{Lyth:1996im}, and for its most generalized updated version in presence of phase velocity at the horizon crossing also see~\cite{Baumann:2011ws}. 
Such a significant tensor to scalar ratio, can be obtained in the framework of large-field models of inflation, such
as ``chaotic inflation'' \cite{Linde:1983gd,Albrecht:1983by}, so-called ``Higgs inflation'' along with its conformal generalization \cite{Nakayama:2010sk,Choudhury:2013zna} and ``axion monodromy inflation'' \cite{Silverstein:2008sg}. In this class of models, slow-roll inflation occurs when the inflaton vacuum
expectation value (VEV) exceeds the Planck scale, so that the large field excursion, $\Delta\phi>M_{p}$ is possible. 
However,  in this paper the main goal is to provide an analytical expression for tensor-to-scalar ratio when $\Delta \phi < M_p$, as suggested in Refs.~\cite{BenDayan:2009kv,Hotchkiss:2011gz}.  As it has been show recently~\cite{Choudhury:2013jya}, it is indeed possible to obtain large $r\leq 0.12$ for field values $\Delta \phi \leq M_p$, here we provide an analytical proof of the earlier results.

Our analytical results are important because any positive detection on large tensor-to-scalar ratio, i.e. $r\sim {\cal O}(0.01-0.1)$, in forthcoming experiments might not be able to conclusively favour high scale super-Planckian models of inflation. 

\section{Generic framework for sub-Planckian inflation}

The tensor to scalar ratio can be  defined by taking into account of
 the higher order corrections, see Refs.~\cite{Easther:2006tv,Stewart:1993bc,Liddle:1994dx}:
\be\label{para 21ea} 
r=16\epsilon_{H}\frac{\left[1-({\cal C}_{E}+1)\epsilon_{H}\right]^{2}}{\left[1-(2{\cal C}_{E}+1)\epsilon_{H}
+{\cal C}_{E}\eta_{H}\right]^{2}}\,,\ee
where ${\cal C}_{E}=4(\ln 2+\gamma_{E})-5$ with $\gamma_{E}=0.5772$ is the {\it Euler-Mascheroni constant} \cite{Easther:2006tv}.
In Eq~(\ref{para 21ea}) the Hubble slow roll parameters $(\epsilon_{H},\eta_{H})$ are defined as:
\begin{eqnarray}\label{hub1}
    \epsilon_{H}=-\frac{d\ln H}{d\ln a}=-\frac{\dot{H}}{H^{2}}\,,~~~~~~~
    \eta_{H}=-\frac{d\ln \dot{\phi}}{d\ln a}=-\frac{\ddot{\phi}}{H\dot{\phi}}\,,
   \end{eqnarray}
where dot denotes time derivative with respect to the physical time.
Now considering  the effect from the leading order dominant contributions from the slow-roll parameters, the Hubble slow-roll parameters
can be expressed in terms of the potential dependent slow-roll parameters, $(\epsilon_{V},~\eta_{V})$, as:
$ \epsilon_{H}\approx\epsilon_{V}+\cdots $, and $ \eta_{H}\approx\eta_{V}-\epsilon_{V}+\cdots $,
where $\cdots$ comes from the higher order contributions of $(\epsilon_{V},~\eta_{V})$. Here the 
slow-roll parameters  $(\epsilon_{V},~\eta_{V})$ are given by in terms of the inflationary potential $V(\phi)$, which can be expressed as:
\begin{eqnarray}\label{ra1}
    \epsilon_{V}=\frac{M^{2}_{p}}{2}\left(\frac{V^{\prime}}{V}\right)^{2}\,,~~~~~~~
    \label{ra2} \eta_{V}={M^{2}_{P}}\left(\frac{V^{\prime\prime}}{V}\right)\,.
   \end{eqnarray}
We would also require two other slow-roll parameters, $(\xi^{2}_{V},\sigma^{3}_{V})$, in our analysis, which are given by:
\begin{eqnarray}\label{ja1}
    \xi^{2}_{V}=M^{4}_{p}\left(\frac{V^{\prime}V^{\prime\prime\prime}}{V^{2}}\right)\,,~~~~~~~~
    \label{ja2} \sigma^{3}_{V}=M^{6}_{p}\left(\frac{V^{\prime 2}V^{\prime\prime\prime\prime}}{V^{3}}\right)\,.
   \end{eqnarray}
With the help above mentioned slow roll parameters, i.e. $\epsilon_{V},~\eta_{V},~\xi_{V}$ and $\sigma_V$, we can recast 
Eq~(\ref{para 21ea}) as:
\be\label{para 21e} 
r\approx16\epsilon_{V}\frac{\left[1-({\cal C}_{E}+1)\epsilon_{V}\right]^{2}}{\left[1-(3{\cal C}_{E}+1)\epsilon_{V}
+{\cal C}_{E}\eta_{V}\right]^{2}}\,
\ee
where we have neglected the contributions from the higher order slow-roll terms, as they are sub-dominant at the leading order.
With the help of 
\be\label{con1}
 \frac{d}{d\ln k}=-M_p\frac{\sqrt{2\epsilon_{H}}}{1-\epsilon_{H}}\frac{d}{d\phi}\,\approx -M_p\frac{\sqrt{2\epsilon_{V}}}{1-\epsilon_{V}}\frac{d}{d\phi}\,,%\\~~~~~~\displaystyle 
\ee
and Eq.~(\ref{para 21e}), we can derive a simple expression for the tensor-to-scalar ratio, $r$, as:
\be\label{con2}
 r=\frac{8}{M^{2}_{p}}\frac{(1-\epsilon_{V})^{2}\left[1-({\cal C}_{E}+1)\epsilon_{V}\right]^{2}}{\left[1-(3{\cal C}_{E}+1)\epsilon_{V}
+{\cal C}_{E}\eta_{V}\right]^{2}}\left(\frac{d\phi}{d{\ln k}}\right)^{2}. 
 \ee
Consequently, we can obtain a new bound on $'r'$  in terms of the momentum scale $(k)$:
\be\begin{array}{llll}\label{con4}
    \displaystyle \int^{{ k}_{cmb}}_{{k}_{e}}\frac{dk}{k}\sqrt{\frac{r({k})}{8}} \\ \displaystyle =
\frac{1}{M_p}\int^{{\phi}_{cmb}}_{{\phi}_{e}}d {\phi}\frac{(1-\epsilon_{V})\left[1-({\cal C}_{E}+1)\epsilon_{V}\right]}{\left[1-(3{\cal C}_{E}+1)\epsilon_{V}
+{\cal C}_{E}\eta_{V}\right]},\\
 \displaystyle \approx \frac{1}{M_p}\int^{{\phi}_{cmb}}_{{\phi}_{e}}d {\phi}(1-\epsilon_{V})\left[1+{\cal C}_{E}(2\epsilon_{V}-\eta_{V})+....\right],\\
\displaystyle \approx \frac{\Delta\phi}{M_p} \left\{ 1+\frac{1}{\Delta\phi}\left[(2{\cal C}_{E}-1)\int^{{\phi}_{cmb}}_{{\phi}_{e}}d {\phi}~\epsilon_{V}
%\right.\right.\\ \left.\left.\displaystyle~~~~~~~~~~~~~~~~~~~~~~~~~~~~~~~
-{\cal C}_{E}\int^{{\phi}_{cmb}}_{{\phi}_{e}}d {\phi}~\eta_{V}\right]+....\right\}\,,
   \end{array}\ee
where note that $\Delta\phi \approx \phi_{cmb}-\phi_{e}$ is positive in Eq.~(\ref{con4}), and this implies the left 
hand side  of the integration over momentum within an interval, $k_{e}<k<k_{cmb}$, is also positive. Individual integrals 
involving $\epsilon_V$ and $\eta_V$ were estimated in an Appendix, see Eqs.~(\ref{hj1}) and (\ref{hj2}).

Here $(\phi_{e},~k_{e})$ and $(\phi_{cmb},~k_{cmb})$ represent inflaton field value and the corresponding momentum scale
at the end of inflation and the Hubble crossing respectively. The imprints of the primordial gravitational waves 
can be directly  measured in the CMB experiments via $r(k_{cmb})$. It is important to note that the
 recent observational constraint from Planck \cite{Ade:2013uln} only fixes the upper bound on  $r(k_{cmb}\approx k_{\star})(\leq 0.12)$ by fixing the upper bound of the 
scale of inflation at the GUT scale $(V_{\star}\lesssim 10^{16}~{\rm GeV})$.

%The lower bound of such constraint is still yet to be determined. If future CMB experiments can detect the signatures of primordial gravitational waves, then it is possible to fix the lower bound of $r(k_{cmb})$ also. See Eq~(\ref{scale}) for the details where the pivot/normalization scale of momentum $k_{\star}$ sets the reference level around which the tensor-to-scalar ratio can be parameterized at any arbitrary momentum scale $k$.

In order to perform the momentum integration in the left hand side of Eq~(\ref{con4}), we have used 
 $r(k)$ at any arbitrary momentum scale, which can be expressed as:
\be\label{con5}
 r(k)=r(k_{\star})\left(\frac{k}{k_{\star}}\right)^{a+\frac{b}{2}\ln\left(\frac{k}{k_{\star}}\right)
+\frac{c}{6}\ln^{2}\left(\frac{k}{k_{\star}}\right)+....}\,,
\ee
where $$a=n_{T}-n_{S}+1,~~~b=\left(\alpha_{T}-\alpha_{S}\right),~~~c=\left(\kappa_{T}-\kappa_{S}\right)$$ are 
explicitly defined in Ref.~\cite{Choudhury:2013jya}. These parameterisation characterises the spectral indices, $n_S,~n_T$, running 
of the spectral indices, $\alpha_S,~\alpha_T$, and running of the running of the spectral indices, $\kappa_S,~\kappa_T$. 
Here the subscript $(S,~T)$ represent the scalar and tensor modes.

It was earlier confirmed by the WMAP9+high-{\it l}+BAO+$H_{0}$ combined constraints that: $\alpha_{S}=-0.023 \pm 0.011$ and $\kappa_{S}=0$ within less than
$1\sigma$ C.L. \cite{Hinshaw:2012aka}. After the Planck release it is important to see the impact on $r(k_\ast)$ due  to running, and running of the running of 
the spectral tilt by modifying the generic power law form of the parameterization of tensor-to-scalar ratio. The combined Planck+WMAP9 constraint 
confirms that: $\alpha_{S}=-0.0134\pm 0.0090$ and $\kappa_{S}=0.020^{+0.016}_{-0.015}$ within $1.5\sigma$ statistical accuracy \cite{Ade:2013uln},
 which additionally includes $\kappa_{S}\neq 0$ possibility.

 At the next to leading order, the simplest way to modify the power law parameterization is to incorporate the effects of 
higher order Logarithmic corrections in terms of the presence of non-negligible running, and running of the running of the spectral tilt as shown in Eq~(\ref{con5}), which involves higher order slow-roll corrections~\footnote{It is important to note that when Ref.~\cite{Lyth:1996im}  first derived a bound on large tensor-to-scalar ratio 
for super-Planckian inflationary models (with $\Delta\phi >M_{p}$), the above mentioned constraints on $\alpha_S, \kappa_S$ were not taken into account due to lack of observational constraints.}.

After substituting Eq~(\ref{con5}) in Eq~(\ref{con4}), we will show that  additional information can be gained from our analysis:
first of all it provides more accurate and improved bound on tensor-to-scalar ratio in presence of non-negligible running and
 running of the running of the  spectral tilt. In our analysis super-Planckian physics doesn't play any role as the effective theory 
puts naturally an upper cut-off set by the Planck scale. Consequently the prescription only holds good for sub-Planckian VEVs, $\phi_{0}<M_{p}$ and field excursion, 
$\Delta\phi<M_{p}$ for inflation. Both these outcomes open a completely new insight into the particle physics
motivated models of inflation, which are valid below the Planck scale.

Further note that the momentum integral has non-monotonous behaviour of the slow-roll parameters ($\epsilon_{V},\eta_{V}$)
 within the interval, $k_{e}<k<k_{cmb}$, which implies that 
$\epsilon_{V}$ and $\eta_{V}$ initially increase within an observable window of e-foldings 
(which we will define in the next section, see Eq~(\ref{efold})), and then decrease at some point during the inflationary epoch when the observable 
scales had left the Hubble patch, and  eventually increase again to end inflation \cite{BenDayan:2009kv,Hotchkiss:2011gz}.

In the most general situation, in Eq.~(\ref{con5}),
the parameters $a,~b$ and $c$ are all functions of arbitrary momentum scale \cite{Choudhury:2013jya}. After imposing the above mentioned
 non-monotonicity behaviour of the slow-roll parameters within this interval, we can easily express the
parameters $a,~b$ and $c$ at the pivot scale $k_{\star}$, which is approximately close to the CMB scale, i.e. $k_{cmb}\approx k_{\star}$.
The computational details of the momentum integration appearing in Eq~(\ref{con4}) are elaborately discussed in the appendix,
see Eq~(\ref{momentumint},~\ref{momentumint1}) and the subsequent discussion. 

Let us now expand a generic inflationary potential around the vicinity of $\phi_0$ where inflation occurs, and impose 
the flatness condition such that, $V^{\prime\prime}(\phi_0)\approx 0$. This yields a potential, see~\cite{Enqvist:2010vd}:
\be\label{rt1a}
V(\phi)=\alpha+\beta(\phi-\phi_{0})+\gamma(\phi-\phi_{0})^{3}+\kappa(\phi-\phi_{0})^{4}+\cdots\,,
\ee
where $\alpha\ll M_p^4$ denotes the height of the potential, 
and the coefficients $\beta \ll M_p^3,~\gamma\ll M_p,~\kappa\ll {\cal O}(1)$ determine the shape of the 
potential in terms of the model parameters. Typically, $\alpha$ can be set to zero by fine tuning, but here we wish to keep this 
term for generality. 

Note that at this point, we do not need to specify any particular model of inflation for Eq.~(\ref{rt1a}). 
However, not all of the coefficients are independent once we prescribe the model of inflation here.This is true {\it only} if the model is fully embedded within a particle theory such as 
that of MSSM~\cite{Allahverdi:2006iq}. We will always observe the crucial constraints: $\phi_0<M_p$ and $\Delta\phi<M_p$.
Then $\Delta\phi$ can be redefined as, 
$\Delta\phi=(\phi_{cmb}-\phi_{0})-(\phi_{e}-\phi_{0})$.

Now substituting the explicit form of the 
potential stated in Eq.~(\ref{rt1a}), we evaluate the crucial integrals of the first and second slow-roll parameters ($\epsilon_{V},~\eta_{V}$)
appearing in the right hand side of Eq.~(\ref{con4}). For details see appendix where the leading order results are explicitly mentioned.

%%%%%%%%%%%%%%%%%%%%%%%%%%%%%%%%%%%%%%%%%%%%%%%%%%%%%%%%%%%%%%%%%

\section{Accurate bound on $'r'$ for small field values of inflation}

At any arbitrary momentum scale the number of
e-foldings, ${\cal N}(k)$, between the Hubble exit of the relevant modes and the end of inflation can be expressed as~\cite{Ade:2013uln}:
%
%\begin{widetext}
\be\begin{array}{llll}\label{efold}
\displaystyle {\cal N}(k) \approx  71.21 - \ln \left(\frac{k}{k_{0}}\right)  
+  \frac{1}{4}\ln{\left( \frac{V_{\star}}{M^4_{P}}\right) }
+  \frac{1}{4}\ln{\left( \frac{V_{\star}}{\rho_{end}}\right) }  
+ \frac{1-3w_{int}}{12(1+w_{int})} 
\ln{\left(\frac{\rho_{rh}}{\rho_{end}} \right)},
\end{array}\ee
%\end{widetext}
%
where $\rho_{end}$ is the energy density at the end of inflation, 
$\rho_{rh}$ is an energy scale during reheating, 
$k_{0}=a_0 H_0$ is the present Hubble scale, 
$V_{\star}$ corresponds to the potential energy when the relevant modes left the Hubble patch 
during inflation corresponding to the momentum scale $k_{\star}\approx k_{cmb}$, and $w_{int}$ characterises the effective equation of state 
parameter between the end of inflation and the energy scale during reheating. 

Within the momentum interval,
$k_{e}<k<k_{cmb}$, the corresponding number of e-foldings is given by, $\Delta {\cal N} = {\cal N}_{e}-{\cal N}_{cmb}$, as
\be\begin{array}{lll}\label{intnk}
    \displaystyle \Delta {\cal N} =\ln\left(\frac{k_{cmb}}{k_{e}}\right)\approx \ln\left(\frac{k_{\star}}{k_{e}}\right)
=\ln\left(\frac{a_{\star}}{a_{e}}\right)+\ln\left(\frac{H_{\star}}{H_{e}}\right)
\approx \ln\left(\frac{a_{\star}}{a_{e}}\right)+\frac{1}{2}\ln\left(\frac{V_{\star}}{V_{e}}\right)\,
   \end{array}\ee
where $(a_{\star},H_{\star})$ and $(a_{e}H_{e})$
represent the scale factor and the Hubble parameter at the pivot scale and end of inflation, and we have used the fact that 
$H^{2} \propto V$. We can estimate the contribution of the last term of the right hand side by using Eq~(\ref{rt1a}) as follows:
\be\begin{array}{llll}\label{espot}
    \displaystyle \ln\left(\frac{V_{\star}}{V_{e}}\right)=\ln\left(
\frac{\alpha+\beta(\phi_{\star}-\phi_{0})+\gamma(\phi_{\star}-\phi_{0})^{3}+\kappa(\phi_{\star}-\phi_{0})^{4}
+\cdots\,}{\alpha+\beta(\phi_{e}-\phi_{0})+\gamma(\phi_{e}-\phi_{0})^{3}+\kappa(\phi_{e}-\phi_{0})^{4}+\cdots\,}\right),\\
\displaystyle ~~~~~~~~~~~\approx\ln\left(1+\underbrace{M_{p}\frac{\beta}{\alpha}\left(\frac{\Delta\phi}{M_{p}}\right)}_{<< 1}[1+\underbrace{\cdots}_{<< 1}]\right),\\
\displaystyle ~~~~~~~~~~~\approx \ln(1+\underbrace{\cdots}_{<< 1}),\\ 
   \end{array}\ee
 where $(\Delta\phi/M_{p})<<1$, and we assume $(\beta M_{p}/\alpha)<<1$, consequently, Eq~(\ref{intnk}) reduces to the following simplified expression:
\be\begin{array}{lll}\label{intnk}
    \displaystyle \Delta {\cal N} \approx \ln\left(\frac{k_{\star}}{k_{e}}\right)
\approx\ln\left(\frac{a_{\star}}{a_{e}}\right) \approx 17~{\rm efolds}\,.
   \end{array}\ee
Within the observed limit of Planck, i.e. $\Delta N\approx 17$, the slow-roll parameters, see  Eqs.~(\ref{hj1},~\ref{hj2}) of Appendix, show non-monotonic behaviour, 
where the corresponding scalar and tensor amplitude of the power spectrum remains almost unchanged~\footnote{In this paper we fix $\Delta {\cal N}\approx 17$ e-foldings as within this interval the combined 
Planck+WMAP9 constraints on the amplitude of power spectrum $\ln(10^{10}P_{S})=3.089^{+0.024}_{-0.027}$ (within $2\sigma$ C.L.), spectral tilt $n_{S}=0.9603 \pm 0.0073$ (within $2\sigma$ C.L.),
 running of the spectral tilt $\alpha_{S}=-0.0134\pm 0.0090$ (within $1.5\sigma$ C.L.) and running of running of spectral tilt
 $\kappa_{S}=0.020^{+0.016}_{-0.015}$ (within $1.5\sigma$ C.L.) are satisfied \cite{Ade:2013uln}.}.
 
 At the scale of Hubble crossing ($k_{\star}=a_{\star}H_{\star}$), the slow-roll parameter $\epsilon_{V}$ must be sufficiently large enough to generate an observable value of tensor-to-scalar ratio $r_{\star}$ at the pivot/normalization scale $k_{\star}$, and it must increase over the $\Delta {\cal N}\approx 17$ e-foldings, as 
 first pointed out in Refs.~\cite{BenDayan:2009kv,Hotchkiss:2011gz}.
After Hubble crossing ($k_{\star}>>a_{\star}H_{\star}$), the slow-roll parameter $\epsilon_{V}$
 must quickly decrease, which is necessary to generate enough e-folds of inflation. However instead of a quick decrement of $\epsilon_{V}$
if it decreases gradually, it will need to eventually decrease to a much smaller value because, $\epsilon_{V}\propto({\Delta\phi}/{M_{p}\Delta {\cal N}})< {1}/{17}$, by imposing the constraint, $\Delta\phi< M_{p}$.

Substituting the results obtained from Eq.~(\ref{hj1}), Eq.~(\ref{hj2}) and Eq~(\ref{momentumint1}) (see Appendix), and with the help of Eq.~(\ref{intnk}), 
up to the leading order, we obtain:
%
%\begin{widetext}
\be\begin{array}{llll}\label{con9}
    \displaystyle \sqrt{\frac{r(k_{\star})}{8}}\left\{\left(\frac{a}{4}-\frac{b}{16}+\frac{c}{48}-\frac{1}{2}\right)\left[1
-\frac{k^{2}_{e}}{k^{2}_{\star}}\right]%+\left(\frac{b}{8}-\frac{c}{24}\right)\left[\frac{k^{2}_{cmb}}{k^{2}_{\star}}
%\ln\left(\frac{k_{cmb}}{k_{\star}}\right)
%\right.\right.\\ \left.\left.\displaystyle~~~~~~~~~~~~~~-\frac{k^{2}_{e}}{k^{2}_{\star}}\ln\left(\frac{k_{e}}{k_{\star}}\right)\right]
%\displaystyle +\frac{c}{24}\left[\frac{k^{2}_{cmb}}{k^{2}_{\star}}
%\ln^{2}\left(\frac{k_{cmb}}{k_{\star}}\right)-\frac{k^{2}_{e}}{k^{2}_{\star}}
%\ln^{2}\left(\frac{k_{e}}{k_{\star}}\right)\right]
+\cdots\,\right\}\\
%\displaystyle = \left\{ \frac{\Delta\phi}{M_p} +\left[\sum^{10}_{m=0} \frac{M^{m+2}_{PL}~{\bf G}_{m}}{(m+1)}\left(\frac{\phi_{e}-\phi_{0}}{M_p}\right)^{m+1}
%\left\{\left(1+\frac{\Delta\phi}{M_p}\left(\frac{\phi_{e}-\phi_{0}}{M_p}\right)^{-1}\right)^{m+1}-
%1\right\}\right]+\cdots\,\right\}
%\\
\displaystyle\approx\left\{ \left(1 +\underbrace{\sum^{10}_{m=0}{\bf A}_{m}
\left(\frac{\phi_{e}-\phi_{0}}{M_p}\right)^{m}}_{\ll 1} \right)\frac{\Delta\phi}{M_p}
%\right.\\ \left. \displaystyle~~~~~~~~~~~~~~~~~~~~~~~~~~~~~
+\underbrace{\sum^{10}_{m=0} \frac{m{\bf A}_{m}}{2}
\left(\frac{\phi_{e}-\phi_{0}}{M_p}\right)^{m-1}
\left(\frac{\Delta\phi}{M_p}\right)^{2}}_{\ll 1}+\cdots\,\right\}\,,
   \end{array}\ee
%\end{widetext}
%
where $(k_{e}/k_{\star})\approx \exp(-\Delta{\cal N})=\exp(-17)\approx 4.13\times 10^{-8}$ and we have defined a new dimensionless 
binomial expansion co-efficient (${\bf A}_{m}$) as:
\be\label{cond}
{\bf A}_{m}=M^{m+2}_{p}\left[\left({\cal C}_{E}-\frac{1}{2}\right){\bf C}_{m}-6{\cal C}_{E}{\bf D}_{m}\right]~~~~(\forall m=0,1,2,....,10)
 \ee
with an additional requirement, ${\bf D}_{m}=0$ for $m=0$ and $m>6$ obtained from the binomial series expansion obtained from the
leading order results of the slow-roll integrals stated in the appendix~\footnote{In Eq.~(\ref{cond}), and 
Eqs.~(\ref{hj1},\ref{hj2}) (see Appendix), ${\bf C}_{p}$ and ${\bf D}_{q}$ are Planck suppressed dimensionful (mass dimension 
$\left[M^{-(m+2)}_{p}\right]$) binomial series expansion co-efficient which are expressed in terms of the generic model 
parameters $(\alpha,\beta,\gamma,\kappa,\cdots)$ as presented
in Eq~(\ref{rt1a}). These co-efficients follow another additional restriction on the indices appearing in the subscript as,
$p=0,1,....,10$, and $q=1,2,....,6$. Instead of using two indices $(p,q)$ if we generalize them by a single index $m$ as mentioned in 
Eq~(\ref{cond}), the above mentioned requirement on ${\bf D}_{m}$ naturally appears in the present context.}.
Additionally it is also important to note that the expansion co-efficient ${\bf A}_{m}(\forall m)$ are suppressed by the various powers of the scale of inflation, 
$\alpha$, which is the leading order term in generic expansion of the inflationary potential as shown in Eq~(\ref{rt1a}) (see Eq~(\ref{ceff}) in the appendix). 
Consequently we can expand the left side of Eq.~(\ref{con9}) 
in the powers of ${\Delta\phi}/{M_p}$, using the additional constraint 
$\Delta\phi<(\phi_{e}-\phi_{0})<M_p$. 
This clearly implies that the highlighted terms by $\underbrace{\cdots}$ are sufficiently smaller than unity for which we can easily neglect the higher order terms of 
${\Delta\phi}/{M_p}$. 

To the  first order approximation - we can take $k_{\star}\approx k_{cmb}$ within  $17$ e-foldings of inflation, and 
neglecting all the higher powers of $k_{e}/k_{\star}\approx {\cal O}(10^{-8})$ from the left hand side of Eq~(\ref{con9}).
Consequently, Eq.~(\ref{con9}) reduces to the following compact form for $r(k_\ast)$:
%
%\begin{widetext}
%\begin{widetext}
\be\begin{array}{llll}\label{con10sd}
    \displaystyle \frac{3}{25}\sqrt{\frac{r(k_{\star})}{0.12}}\left|\left\{\frac{3}{400}\left(\frac{r(k_{\star})}{0.12}\right)-\frac{\eta_{V}(k_{\star})}{2}-\frac{1}{2}%\right]
-\left(6{\cal C}_{E}+\frac{14}{3}\right)\epsilon^{2}_{V}(k_{\star})
 -\frac{\eta^{2}_{V}(k_{\star})}{6}\right.\right.\\ \left.\left.~~~~~~~~~~
\displaystyle +\left({\cal C}_{E}-\frac{1}{4}\right)\frac{\xi^{2}_{V}(k_{\star})}{2}-\left(2{\cal C}_{E}-\frac{5}{12}\right)\eta_{V}(k_{\star})\epsilon_{V}(k_{\star})-\frac{\sigma^{3}_{V}(k_{\star})}{24}+\cdots\,\right\}\right|
\displaystyle \approx \frac{\left |\Delta\phi\right|}{M_p}\,,
   \end{array}\ee
%\end{widetext}
%\end{widetext}
%
provided at the pivot scale, $k=k_{\star}\approx k_{cmb}>>k_{e}$, here $\eta_{V}>>\left\{\epsilon^{2}_{V},\eta^{2}_{V},\xi^{2}_{V},\sigma^{3}_{V},\cdots\right\}$
approximation is valid for which at the leading order, the first three terms dominate 
over the other higher order contributions appearing in the right hand side of Eq~(\ref{con10sd}). 

Now, it is also possible to recast $a(k),~b(k),~c(k)$, in terms of $r(k)$, and the slow roll parameters by using the 
relation, Eq.~(\ref{para 21e}), to write:
\be\begin{array}{lll}\label{apara}
    \displaystyle a(k_{\star})\approx \left[\frac{r(k_{\star})}{4}-2\eta_{V}(k_{\star})-4\left(2{\cal C}_{E}+\frac{1}{3}\right)\epsilon_{V}(k_{\star})\eta_{V}(k_{\star})
-4\left(6{\cal C}_{E}+\frac{11}{3}\right)\epsilon^{2}_{V}(k_{\star})
\right.\\ \left.\displaystyle~~~~~~~~~~~~~~~~~~~~~~~~~~~~~~~~~~~~~~~~~~~~~~~~~~~~~~~~~~~+2{\cal C}_{E}\xi^{2}_{V}(k_{\star})-\frac{2}{3}\eta^{2}_{V}(k_{\star})+\cdots\right],\\
    \displaystyle b(k_{\star})\approx \left[16\epsilon^{2}_{V}(k_{\star})-12\epsilon_{V}(k_{\star})\eta_{V}(k_{\star})+2\xi^{2}_{V}(k_{\star})+\cdots\right],\\
\displaystyle c(k_{\star})\approx \left[-2\sigma^{3}_{V}+\cdots\right],
   \end{array}\ee
where $``\cdots''$ involves higher order slow-roll contributions which are negligibly small in the leading order approximation.
The additional constraint $a>>b>>c$  defined in Eq.~(\ref{con5}) is always 
satisfied by the general class of inflationary potentials,  for instance the saddle or the inflection
 point models of inflation do satisfy this constraint~\cite{Allahverdi:2006iq}.

The recent observations from {\it Planck} puts an upper bound on gravity waves
via  tensor-to-scalar ratio as $r(k_{\star})\leq 0.12$ at the pivot scale, $k_{\star}=0.002~Mpc^{-1}$~\cite{Ade:2013uln}:
\begin{equation}\label{scale}
     V_{\star}\leq (1.96\times 10^{16}{\rm GeV})^{4}~\frac{r(k_{\star})}{0.12}.
   \end{equation}
 Combining Eqs.~(\ref{con10sd}) and (\ref{scale}),
we have obtained a closed relationship between $V_*$ and $\Delta \phi$, as:
\be\begin{array}{llll}\label{con12}
\displaystyle \frac{\left|\Delta\phi\right|}{M_p}  \leq  \displaystyle \frac{\sqrt{V_{\star}}}{(2.20\times 10^{-2}~M_p)^{2}}\left|\left\{\frac{V_{\star}}{(2.78\times 10^{-2}M_p)^{4}}-\frac{\eta_{V}(k_{\star})}{2}-\frac{1}{2}%\right]
-\left(6{\cal C}_{E}+\frac{14}{3}\right)\epsilon^{2}_{V}(k_{\star})
 \right.\right.\\ \left.\left.~~~~~~~
~\displaystyle -\frac{\eta^{2}_{V}(k_{\star})}{6}+\left({\cal C}_{E}-\frac{1}{4}\right)\frac{\xi^{2}_{V}(k_{\star})}{2}-\left(2{\cal C}_{E}-\frac{5}{12}\right)\eta_{V}(k_{\star})\epsilon_{V}(k_{\star})-\frac{\sigma^{3}_{V}(k_{\star})}{24}+\cdots\,\right\}\right|\,,
   \end{array}\ee
where
$\eta_{V}>>\left\{\epsilon^{2}_{V},\eta^{2}_{V},\xi^{2}_{V},\sigma^{3}_{V},\cdots\right\}$ are satisfied, and at the leading order first three terms dominate 
over the other higher order contributions, therefore 
\begin{equation}\label{con13}
\frac{|\Delta\phi|}{M_p}  \leq  \frac{\sqrt{V_{\star}}}{(2.20\times 10^{-2}~M_p)^{2}}\left|\frac{V_{\star}}{(2.78\times 10^{-2}M_p)^{4}}-\frac{\eta_{V}(k_{\star})}{2}-\frac{1}{2} \right|\,.
\end{equation}
The above Eqs.~(\ref{con12},~\ref{con13}) are new improved bounds on $\Delta \phi$ during a slowly rolling single field  $\phi$ within an effective 
field theory treatment, where the vev of an inflaton remains sub-Planckian, i.e. $\phi_0 < M_p$ and $\Delta \phi \ll M_p$. From Eq.~(\ref{con10sd}),
 we can see that large $r\sim 0.1$ can be obtained for models of inflation where
 inflation occurs below the Planck cut-off. Our conditions, Eqs.~(\ref{con10sd},~\ref{con12}), provide 
new constraints on model building for inflation within  particle theory, where the inflaton potential is always constructed 
within an effective field theory with a cut-off. Note that $\eta_V(k_*)\geq 0 $ can provide the largest contribution, 
in order to satisfy the current bound on $r\leq 0.12$,
the shape of the potential has to be {\it concave}.

 \section{Summary and Discussion}

To summarize, in this paper we have presented an accurate bound on tensor to scalar ratio, $r$, 
 and $\Delta\phi$ for a sub-Planckian models of inflation in presence of higher oder slow-roll correction, 
 see Eqs.~(\ref{con10sd},~\ref{con12},~\ref{con13}). The bounds obtained here satisfy the numerical estimations made for
 inflation models based on saddle or inflection points with sub-Planckian VEVs.
 
Further, we have shown that it is indeed possible to realize large tensor-to-scalar ratio
 for sub-Planckian vevs of inflation by assuming the non-monotonicity of the slow-roll parameter $\epsilon_{V}$. Our constraints would help inflationary
model builders and perhaps would enable us to reconstruct the inflationary
 potential \cite{Peiris:2006ug,Cline:2006db,Liddle:1998td,Turner:1995ge} for a single field model of inflation.
 We have also analyzed the fact that the additional constraint on slow-roll parameter, $\eta_V(k_*)\geq 0 $, at the pivot scale of momentum, $k_{\star}$,
also restricts the shape of the potential to be a {\it concave} one.

%%%%%%%%%%%%%%%%%%%%%%%%%%%%%%%%%%%%%%%%%%%%%%%%%%%%%%%%%%%%%%%%%%%%%%%%%%%%%%%%%%%%%%%%%%%%%%%%%%%%%%%%%%%%%%%%%%%%%%%%%%%%%%%%%%%%%%%%%%%%%%%%%%%%%%%%%%%%%%%%%%%%%%%%%%%%%%%%%%%%%%%%%%%%%%%%%%%%%%%%%%%%%%%%%%%%%%%%%%%%%
%%%%%%%%%%%%%%%%%%%%%%%%%%%%%%%%%%%%%%%%%%%%%%%%%%%%%%%%%%%%%%%%%%%%%%%%%%%%%%%%%%%%%%%%%%%%%%%%%%%%%%%%%%%%%%%%%%%%%%%%%%%%%%%%%%%%%%%%%%%%%%%%%%%%%%%%%%%%%%%%%%%%%%%%%%%%%%%%%%%%%%%%%%%%%%%%%%%%%%%%

\section*{Acknowledgments}
SC thanks Council of Scientific and
Industrial Research, India for financial support through Senior
Research Fellowship (Grant No. 09/093(0132)/2010). AM is supported 
by the Lancaster-Manchester-Sheffield Consortium for Fundamental Physics under STFC grant ST/J000418/1.

%%%%%%%%%%%%%%%%%%%%%%%%%%%%%%%%%%%%%%%%%%%%%%%%%%%%%%%%%%%%%%%%%%%%%%%%%%%%%%%%%%%%%%%%%%%%%%%%%%%%%%%%%%%%%%%%%%%%%%%%%%%%%%%%%%%%%%%%%%%%%%%%%%%%%%%%%%%%%%%%%%%%%%%%%%%%%%%%%%%%%%%%
%%%%%%%%%%%%%%%%%%%%%%%%%%%%%%%%%%%%%%%%%%%%%%%%%%%%%%%%%%%%%%%%%%%%%%%%%%%%%%%%%%%%%%%%%%%%%%%%%%%%%%%%%%%%%%%%%%%%%%%%%%%%%%%%%%%%%%%%%%%%%%%%%%%%%%%%%%%%%%%%%%%%%%%%%%%%%%%%%%%%%%%%%%%%%%%%%%%
\section*{Appendix}

{\Large \bf A. Slow-roll integration}\\ \\
The crucial integrals of the first and second slow-roll parameters ($\epsilon_{V},~\eta_{V}$)
appearing in the right hand side of Eq.~(\ref{con4}),
 which can be written up to the leading order as: 
%
%\begin{widetext}
\be\begin{array}{llll}\label{hj1}
    \displaystyle \int^{{\phi}_{cmb}}_{{\phi}_{e}}d {\phi}~\epsilon_{V}=\frac{M^{2}_{p}}{2}\int^{{\phi}_{cmb}}_{{\phi}_{e}}d {\phi}~
\frac{\left[\beta+3\gamma(\phi-\phi_{0})^{2}+4\kappa(\phi-\phi_{0})^{3}+\cdots\,\right]^{2}}{\left[\alpha
+\beta(\phi-\phi_{0})+\gamma(\phi-\phi_{0})^{3}+\kappa(\phi-\phi_{0})^{4}+\cdots\,\right]^{2}},\\
\displaystyle ~~~~~~~~~~~~~~~~\approx\frac{1}{2}\sum^{10}_{p=0} \frac{M^{p+2}_{p}~{\bf C}_{p}}{(p+1)}\left(\frac{\phi_{e}-\phi_{0}}{M_p}\right)^{p+1}
\left\{\left(1+\frac{\Delta\phi}{M_p}\left(\frac{\phi_{e}-\phi_{0}}{M_p}\right)^{-1}\right)^{p+1}-
1\right \}+\cdots\,
   \end{array}\ee
 \be\begin{array}{llll}\label{hj2}
    \displaystyle \int^{{\phi}_{cmb}}_{{\phi}_{e}}d {\phi}~\eta_{V}=6M^{2}_{p}\int^{{\phi}_{cmb}}_{{\phi}_{e}}d {\phi}~
\frac{\left[\gamma(\phi-\phi_{0})+2\kappa(\phi-\phi_{0})^{2}+\cdots\,\right]}{\left[\alpha
+\beta(\phi-\phi_{0})+\gamma(\phi-\phi_{0})^{3}+\kappa(\phi-\phi_{0})^{4}+\cdots\,\right]},\\
\displaystyle ~~~~~~~~~~~~~~~~\approx 6\sum^{6}_{q=1} \frac{M^{q+2}_{p}~{\bf D}_{q}}{(q+1)}\left(\frac{\phi_{e}-\phi_{0}}{M_p}\right)^{q+1}
\left\{\left(1+\frac{\Delta\phi}{M_p}\left(\frac{\phi_{e}-\phi_{0}}{M_p}\right)^{-1}\right)^{q+1}-
1\right \}+\cdots\,
   \end{array}\ee
% \end{widetext}
%
where we have used the  $(\phi-\phi_{0})<M_p$  (including at $\phi=\phi_{cmb}$ and $\phi=\phi_{e}$) around $\phi_0$.
The leading order dimensionful Planck scale suppressed expansion co-efficients (${\bf C}_{p}$) and (${\bf D}_{q}$)
 are given in terms of the model parameters $(\alpha,\beta,\gamma,\kappa)$, which determine the
hight and shape of the potential in terms of the model parameters as:
%
%\begin{widetext}
\be\begin{array}{llll}\label{ceff}
    \displaystyle {\bf C}_{0}=\frac{\beta^{2}}{\alpha^{2}},~~{\bf C}_{1}=-\frac{2\beta^{3}}{\alpha^{3}},~~{\bf C}_{2}=\frac{6\beta\gamma}{\alpha^{2}},~~
{\bf C}_{3}=\frac{8\beta\kappa}{\alpha^{2}}-\frac{14\beta^{2}\gamma}{\alpha^{3}},~~{\bf C}_{4}=\frac{9\gamma^{2}}{\alpha^{2}}-\frac{18\beta^{2}\kappa}{\alpha^{3}},\\%~~
\displaystyle {\bf C}_{5}=\frac{24\gamma\kappa}{\alpha^{2}}-\frac{30\gamma^{2}\beta}{\alpha^{3}}-\frac{48\gamma^{2}\beta\kappa}{\alpha^{3}},%\\
\displaystyle ~~~~{\bf C}_{6}=\frac{16\kappa^{2}}{\alpha^{2}}-\frac{32\kappa^{2}\beta}{\alpha^{3}}-\frac{28\beta\gamma\kappa}{\alpha^{3}},\\%~~
\displaystyle {\bf C}_{7}=-\frac{18\gamma^{3}}{\alpha^{3}}-\frac{16\beta\kappa^{2}}{\alpha^{3}},~~
{\bf C}_{8}=-\frac{66\gamma^{2}\kappa}{\alpha^{3}},~~{\bf C}_{9}=-\frac{80\gamma\kappa^{2}}{\alpha^{3}},~~{\bf C}_{10}=-\frac{32\kappa^{3}}{\alpha^{3}},\\
    \displaystyle {\bf D}_{1}=\frac{\gamma}{\alpha},~~{\bf D}_{2}=\frac{2\kappa}{\alpha}-\frac{\beta\gamma}{\alpha^{2}},~~{\bf D}_{3}=-\frac{2\kappa\beta}{\alpha^{2}},~~
{\bf D}_{4}=-\frac{\gamma^{2}}{\alpha^{2}},~~{\bf D}_{5}=-\frac{3\kappa\gamma}{\alpha^{2}},~~{\bf D}_{6}=-\frac{2\kappa^{2}}{\alpha^{2}}.
   \end{array}\ee
\\ \\
{\Large \bf B. Momentum integration}\\ \\
The momentum integral appearing in the left hand side of the Eq~(\ref{con4}) is computed as:
\be\begin{array}{llll}\label{momentumint}
    \displaystyle \int^{{ k}_{cmb}}_{{k}_{e}}\frac{dk}{k}\sqrt{\frac{r({k})}{8}}\\
\displaystyle =\sqrt{\frac{r(k_{\star})}{8}}\int^{{ k}_{cmb}}_{{k}_{e}}\frac{dk}{k}\left(\frac{k}{k_{\star}}\right)^{a+\frac{b}{2}\ln\left(\frac{k}{k_{\star}}\right)
+\frac{c}{6}\ln^{2}\left(\frac{k}{k_{\star}}\right)+....}\, \\ 
\displaystyle =\left\{\begin{array}{ll}
                    \displaystyle  \sqrt{\frac{r(k_{\star})}{8}}\left\{\left(\frac{a}{4}-\frac{b}{16}+\frac{c}{48}-\frac{1}{2}\right)\left[\frac{k^{2}_{cmb}}{k^{2}_{\star}}
-\frac{k^{2}_{e}}{k^{2}_{\star}}\right]\right.\\ \left. 
\displaystyle~~~~~~~~~ +\left(\frac{b}{8}-\frac{c}{24}\right)\left[\frac{k^{2}_{cmb}}{k^{2}_{\star}}
\ln\left(\frac{k_{cmb}}{k_{\star}}\right)-\frac{k^{2}_{e}}{k^{2}_{\star}}\ln\left(\frac{k_{e}}{k_{\star}}\right)\right]
\right.\\ \left.\displaystyle~~~~~~~~~~ +\frac{c}{24}\left[\frac{k^{2}_{cmb}}{k^{2}_{\star}}
\ln^{2}\left(\frac{k_{cmb}}{k_{\star}}\right)-\frac{k^{2}_{e}}{k^{2}_{\star}}
\ln^{2}\left(\frac{k_{e}}{k_{\star}}\right)\right]+\cdots\,\right\} &
 \mbox{ {\bf for} $a,b,c \neq 0$ \& $a>b>c$}  \\ \\
         \displaystyle  \sqrt{\frac{r(k_{\star})}{8}}\left\{\left(\frac{a}{4}-\frac{b}{16}-\frac{1}{2}\right)\left[\frac{k^{2}_{cmb}}{k^{2}_{\star}}
-\frac{k^{2}_{e}}{k^{2}_{\star}}\right]\right.\\ \left. 
\displaystyle~~~~~~~~~ +\frac{b}{8}\left[\frac{k^{2}_{cmb}}{k^{2}_{\star}}
\ln\left(\frac{k_{cmb}}{k_{\star}}\right)-\frac{k^{2}_{e}}{k^{2}_{\star}}\ln\left(\frac{k_{e}}{k_{\star}}
\right)\right]+\cdots\,\right\} & \mbox{ {\bf for} $a,b\neq0$, $c=0$ \& $a>b$}\\ \\
\displaystyle  \frac{1}{a+1}\sqrt{\frac{r(k_{\star})}{8}}\left[\left(\frac{k_{cmb}}{k_{\star}}\right)^{a}
-\left(\frac{k_{e}}{k_{\star}}\right)^{a}\right] & \mbox{ {\bf for} $a\neq0$ \& $b,c=0$}\\ \\
\displaystyle  \sqrt{\frac{r(k_{\star})}{8}}\ln\left(\frac{k_{cmb}}{k_{e}}\right) & \mbox{ {\bf for} $a,b,c=0$}.
          \end{array}
\right.
   \end{array}\ee

Further using $k_{cmb}\approx k_{\star}$ and $(k_{e}/k_{\star})\approx \exp(-\Delta{\cal N})=\exp(-17)\approx 4.13\times 10^{-8}$, within 17 e-foldings
Eq~(\ref{momentumint}) can be simplified to the following expression:
\be\begin{array}{llll}\label{momentumint1}
    \displaystyle \int^{{ k}_{cmb}}_{{k}_{e}}\frac{dk}{k}\sqrt{\frac{r({k})}{8}}
\approx\displaystyle\left\{\begin{array}{ll}
                    \displaystyle  \sqrt{\frac{r(k_{\star})}{8}}\left(\frac{a}{4}-\frac{b}{16}+\frac{c}{48}-\frac{1}{2}\right)+\cdots &
 \mbox{ {\bf for} $a,b,c \neq 0$ \& $a>b>c$}  \\ 
         \displaystyle  \sqrt{\frac{r(k_{\star})}{8}}\left(\frac{a}{4}-\frac{b}{16}-\frac{1}{2}\right)+\cdots & \mbox{ {\bf for} $a,b\neq0$, $c=0$ \& $a>b$}\\ 
\displaystyle  \frac{1}{a+1}\sqrt{\frac{r(k_{\star})}{8}}+\cdots & \mbox{ {\bf for} $a\neq0$ \& $b,c=0$}\\ 
\displaystyle  {\cal O}(17)\times\sqrt{\frac{r(k_{\star})}{8}} & \mbox{ {\bf for} $a,b,c=0$}.
          \end{array}
\right.\end{array}\ee

where the last possibility $a,~b,~c=0$ surmounts to the Harrison \& Zeldovich spectrum, which 
is completely ruled out by Planck+WMAP9 data within $5\sigma$ C.L.
Similarly the next to last possibility $a\neq0$ \& $b,~c=0$ is also tightly constrained by the WMAP9 and Planck+WMAP9 data within $2\sigma$ C.L.

The second possibility $a,~b\neq0$, $c=0$, and $a>b$ is favoured by WMAP9 data and tightly constrained within $2\sigma$ window by Planck+WMAP9 data.

Finally, $a,b,c \neq 0$, and  $a>b>c$ case is satisisfied by both WMAP9 and Planck+WMAP9 data within $2\sigma$ C.L. 
Here $a>b>c$ is the only critera which is always satisfied by a general class of inflationary potentials. In this article,  we have only focused on the first 
possibility, i.e. $a>b>c$, from which we have derived all the constraint conditions for a generic model of sub-Planckian inflationary potentials.

%%%%%%%%%%%%%%%%%%%%%%%%%%%%%%%%%%%%%%%%%%%%%%%%%%%%%%%%%%%%%%%%%%%%%%%%%%%%%%%%%%%%%%%%%%%%%%%%%%%%%%%%%%%%%%%%%%%%%%%%%%%%%%%%%%%%%%%%%%%%%%%%%%%%%%%%%%%%%%%%%%%%%%%%%%%%%%%%%%%%%%%%%%%
%%%%%%%%%%%%%%%%%%%%%%%%%%%%%%%%%%%%%%%%%%%%%%%%%%%%%%%%%%%%%%%%%%%%%%%%%%%%%%%%%%%%%%%%%%%%%%%%%%%%%%%%%%%%%%%%%%%%%%%%%%%%%%%%%%%%%%%%%%%%%%%%%%%%%%%%%%%%%%%%%%%%%%%%%%%

\end{document}